# Bandwidth in the Cloud

José Luis García-Dorado,
Universidad Tecnica del Norte, Ecuador.

Wednesday 5$^{\text{th}}$ August, 2015

**Abstract**

The seek for the best quality of service has led Cloud infrastructure clients to disseminate their services, contents and data over multiple cloud data-centers often involving several Cloud Service Providers (CSPs). The consequence of this is that a large amount of bytes must be transmitted across the public Cloud. However, very little is known about its bandwidth dynamics. To address this, we have conducted a measurement campaign for bandwidth between eighteen data-centers of four major CSPs. Such extensive campaign allowed us to characterize the resulting time series of bandwidth as the addition of a stationary component and some infrequent excursions (typically, downtimes). While the former provides a description of the bandwidth users can expect in the Cloud, the latter is closely related to the robustness of the Cloud (i.e., the occurrence of downtimes is correlated). Both components have been studied further by applying a factor analysis, specifically ANOVA, as a mechanism to formally compare data-centers' behaviors and extract generalities. The results show that the stationary process is closely related to data-center locations and CSPs involved in transfers, which fortunately makes both the Cloud more predictable and the set of reported measurements extrapolate. On the other hand, although the correlation in the Cloud is low, i.e., only 10% of the measured pair of paths showed some correlation, we have found evidence that such correlation depends on the particular relationships between pairs of data-centers with little link to more general factors. Positively, this implies that data-centers either at the same area or CSP do not show qualitatively more correlation than others data-centers, which eases the deployment of robust infrastructures. On the downside, this metric is barely generalizable and, consequently, calls for exhaustive monitoring.

## 1 Introduction

To provide the best quality of service, Cloud Service Providers (CSPs), e.g., Amazon EC2, Microsoft Azure, Rackspace or Google Cloud, have pointed to the dispersion of their data-centers across the world. Such geo-distribution ensures both high availability and reliability and also reduces the final users' latency, given its significant impact in business revenue. For example, Amazon estimated that every 100 ms of latency costs 1% in sales [17].

There are numerous examples of research and industry efforts that exploit this paradigm. Netflix is a significant example of application that disseminates its contents over a CSP infrastructure to successfully provide on-demand media services. Geo-distribution has also become a popular and safe way to backup data. Dropbox or Google Drive as well as bank networks are good examples of this. Similarly, Cassandra [20] or RON [2] applications deploy overlay and distributed architectures in the Cloud to take shelter from both network failures and congestion. In addition to these examples, software distributions (such as popular operating system releases), cloud storage, distributed databases [7], virtual machines clones [32] also share the task of moving a large amount of that data to disseminate it over different data-centers [23].

Importantly, while in the past the set of data-centers where a cloud clients deployed a service or application typically belonged to the same CSP, nowadays this scenario tend to be less often. The deployment of both applications and services over different CSPs has proven to be a fundamental tool to provide the lowest latencies [33] and enhanced robustness [27] in the Cloud. Essentially, the limitations a particular CSP can present, e.g. spatially (a poorly covered geographical area) or temporally (a period of malfunction), can be compensated by others, then, making multi-CSP deployments a better approach than their single-CSP counterparts.

To give some representative figures of the importance of traffic between point of presences, some Internet use surveys [12] have shown that more than 77% data-centers operators run backup and replication applications among three or more sites, while more than 50% report over a PB of data in their primary data-center. Further, 70% of surveyed IT firms had between 1 and 10 Gb/s running between data-centers, nearly half having 5 Gb/s or more (between 330GB and 3.3PB per month).

However, while there has been much effort in the research community directed at studying cost-effective bulk data transfers over data-centers in the public Cloud [10, 24, 13], we highlight that very little is known about its bandwidth. Especially, we refer to general results aiming at the identification of invariants on this measure [11]. We believe that such lack of studies is due to the difficulties in measuring traffic from a large number of data-centers on several CSPs over a significant period of time. This paper aims at filling this gap.

We deployed a multi-point of presence testbed that includes eighteen data-centers spread over four major CSPs, and measured the TCP bandwidth of all the paths between them. We noticed that bandwidth in the Cloud can be modeled by a principal Gaussian component and some infrequent excursions. We have related such two behaviors to a stationary process that represents the typical state of a path, and a peak/downtime process that represents the times that a path behaves unusually.

The characterization of the stationary state allows us to answer questions related to what a user can expect of the Cloud performance. While, the study of non-usual behaviors allows us to dig into the robustness of the Cloud. That is, a bad performance on one data-center can be alleviated with, at least, a regular behavior from another. In other words, if the Cloud is correlated.

We apply a factor analysis on these two components with factors such as the time of the day, day of the week, the geographical area data-centers are located, the CSP they belong, and the specific data-centers involved in the measurement of a path. While the latter factor accounts for the peculiarities each data-center or pair of them can have on its behavior, the others will allow us to find generalities to explain the phenomenon. For example, how much of the bandwidth can be explained because a data-center belongs to a given CSP, because a pair of data-centers belong to the same CSP, or depending on the time of the day the sample was gathered. Similarly, when we study downtimes such factor analysis will show if data-centers either in the same area or CSP change their performance in unison, or, otherwise, performance changes in data-centers individually. This exerts a clear impact on how to geo-distribute a deployment in the Cloud. For example, should all data-centers decrease their performance at same time the availability would be in risk, but whether only data-centers' performance of a specific CSP or area dip simultaneously, reliability can be found in other CSP or areas, respectively.

In particular, we have applied the Analysis of Variance (ANOVA) as factor analysis. Interestingly, it has shown that the stationary behavior of a path between data-centers in the Cloud depends strongly on the CSP and areas of the pair (source/destination) of data-centers involved, and qualitatively much less on other factors such as the particular pair of data-centers involved. This supports the generalization and extrapolation of the results herein shown, and even the use of smarter monitoring systems [34]. Similarly, other factors as the time of the day and the day of the week only showed moderate significance. That is, the Cloud is mostly insensitive to the time with some more capacity during weekends, conclusion with a clear impact on the scheduling of bulk transfers in the Cloud.

On the other hand, we found no evidence of additional correlation of the bandwidth time series within areas or CSPs. In general, time series are weakly correlate and such correlation is only marginal explained by location and CSP factors. In this way, the correlation exhibited by some paths is mostly the result of particularities of the data-centers involved. This points at simpler Cloud deployments, as data-centers in the same area or CSP may contribute in the search for availability and reliability equivalently to distant nodes of other CSPs. Unfortunately, this also implies that the generalization of this metric is challenging as similar data-centers behave differently, which calls for an exhaustive and fine-grain monitoring.

We believe that the light shed on the bandwidth dynamics in the Cloud results useful for CSPs as it provides them with a fair comparative description of their performances. Also for practitioners as they can find the description of the regular performance of the Cloud useful for their current deployment and the factorial study useful for future ones. For example, as parameter inputs for novel Cloud simulators [5]. Similarly, the conclusions about correlation can them to make better decisions in terms of both robustness and cost. Finally, for the research community, we believe that this work represents a further step in the path of characterizing the Internet, a task initiated by institutions such as RIPE or Caida [6] years ago. On our part, we focus specifically on an important fraction of the Internet, the public Cloud traffic, remarking on invariants that potentially lead to new models and ideas.

The rest of this work is organized as follows: Section 2 presents the foundations of the paper, this includes our testbed, how to measure bandwidth time series in the Cloud, and their modeling. Sections 3 and 4 present the core results of this work where the stationary and downtimes processes are both described and discussed. Section 5 is devoted to the revision of the related work, and, finally, Section 6 concludes this paper by remarking on the main conclusions and pointing out some lines of future work.

## 2 Preliminaries

The goal of this section is to achieve a comprehensive characterization of the Cloud bandwidth and its interactions. In this light, the first two practical questions we pose are, first, how much time a path between two data-centers must be measured to obtain a significant sample of bandwidth; and, second, on how to formally compare a number of bandwidth time series paths. Let us in first place explain the testbed used to answer them.

### 2.1 Testbed

We have borrowed the approach followed in [22] and [16] where some virtual machines (VM) were started and measurements were gathered between them. Specifically, the authors in [22] studied four CSPs, namely, Amazon EC2, Microsoft Azure, Rackspace Cloud and Google Cloud. As of 2010, such set of CSPs comprised the two most popular ones and two promising newcomers. Currently, they have become in the four dominate players in the Cloud arena [1]. In such a work, the authors referred to CSPs as $C1 \cdots C4$ instead of using their names as an attempt to keep the focus on the conclusions rather than on very specific values. Following a similar approach, we study these same four major CSPs, and use an equivalent name terminology.

Specifically, in this set of CSPs, we started a VM in each of their eighteen data-centers available by the beginning of 2015. Fig. 1 places such data-centers according to the available information CSP provides (some with accurate coordinates, and others more vague ones). For its part, the Table 1 details different features (CSP, geographical area and data-center name) for each of them. In what follows, such features will be used as explanatory factors to the bandwidth phenomenon in the Cloud.

To monitor the bandwidth between data-centers, we developed *CloudB*. It is a scheduler and wrapper of other testing tools, in particular, it executes a set of provided tools with configurable duration and frequency (or under-demand operation) between a list of IP addresses received as input. The tool chosen to measure the bandwidth was Iperf [28], a well-tested software to measure the TCP bandwidth between two nodes. Note that Iperf measures the TCP bandwidth not the maximum capacity of the infrastructure [18]. This ties in with our approach, to unveil the capacity a user can effectively achieve in the Cloud.

In measuring TCP bandwidth in a wide-are a sense, both TCP flow-control and VM networking capacity can act as undesirable bottlenecks to the goal of measuring the Cloud capacity itself. Both [22] and [16] studied to the former. The send and receive windows size are set to 16 MB as larger window sizes did not result in higher measurements. Our initial tests confirmed this threshold as satisfactory.

The later bottleneck has not received so much attention as bandwidth outside data-centers' infrastructure was supposed to be lower 1 Gb/s (often the VM-interface dedicated capacity for medium or large instances). Some of our initial tests suggested that such assumption was valid, and, consequently, the VMs on our testbed were configured with at least 1 Gb/s of DEDICATED network capacity. The specific names for the flavors of VM used vary from one CSP to another, for further details we refer to CSPs' datasheets. Otherwise the total budget reached prohibitive figures as allocating better VMs represent a significant cost increasement. Actually, the measurements gathered throughout this paper have already cost several thousands dollars.

However it is worth remarking that in such tests a few samples of a few paths were relatively closer to 1 Gb/s, i.e., several dozens of Mb/s below 1 Gb/s. To dig into this, we replaced 1 Gb/s-equiped VMs at these data-centers with high-performance VMs flavors. The difference in the measured bandwidth were marginal (even, slightly better in the cheapest ones [16]). Later the full measurement campaign (Section 3.1) confirmed this point, suggesting that our testbed is adequate for our purposes and our concerns were excessive.

## 2.2 Measuring the bandwidth of a path

As TCP bandwidth oscillates with time, we focus on the duration of the process of gathering a bandwidth sample. In our extensive testbed, we measured the TCP bandwidth between 100 randomly-picked paths. The duration was 15 minutes with 1-second granularity. Fig. 2 shows the results for four representative behaviors with the 95% confidence intervals for the mean. They are representative in terms of spanning diverse coefficient of variations (CV) [8] for the bandwidth time series. The figure depicts CVs ranging from 0.09 to 0.46 (minimum and maximum values) and two intermediate examples. Intuitively, the larger the CV is, the duration of the measurement has to be aggregated. Fortunately, Fig. 3 proves that CVs tend to be small. Essentially, its empirical cumulative distribution function (ECDF) shows that all samples are below the rule-of-thumb threshold of 0.5 for significance of the mean, and more than 80% of the paths below 0.2. This indicates that most of the paths behaviors resemble the first examples.

Let see how this translates to time. Note that we are seeking for the smallest interval of time (inter-Cloud bandwidth is not cheap) such that it still provides a significant sample of the bandwidth a transfer achieves in the Cloud. By a transfer in the Cloud, we refer to tasks as those carried out by multimedia content distribution, backup and replication, or distributed storage as the previous section remarked. In this light, to estimate the error that a short measurement makes, we consider as the ground-truth bandwidth of a sample its average throughput after 15 minutes of aggregation. Note that this could represent a reasonable download of 100-10 GB for 1 Gb/s-100 Mb/s rate, which ties in with our scenario.

In this way, Fig. 4 shows the error ratio per path as the difference between the average throughput after $N$ seconds of aggregation (horizontal axis) and the average after 15 minutes. We use a progressive gray-scale where ratios larger than 0.25 are completely black colored. Assuming a ratio error threshold of 0.1-0.05, we note that more than 50% of paths needed more than one minute worth of data. Roughly 20% of paths more than 200 seconds, and after 300 seconds only 5% of samples still show ratios over 0.1 but, in any case, below 0.15. As a trade-off between time aggregation and cost, we deem 5 minutes to be a good compromise.

## 2.3 Modeling paths' behavior

To tackle the modeling of the bandwidth time series, we measured 100 paths for three working days every other 15 minutes for 300 seconds, which translates into 288 samples in each time series. Fig. 5 illustrates with representative examples four different patterns of behavior we found. Specifically, it shows the empirical probability density function for the samples that make up the bandwidth time series (after a Gaussian kernel softening process [4]).

The common denominator of all of them is a clear main component, fairly Gaussian [29] with slightly negative skewness. Roughly 60% of the paths do not present any other component as in the first case of the figure. The other three cases illustrate a progressive increase on the negative skewness (cases II and II) and the last one (IV), additionally, certain probability in the tail. We classify evenly the remaining 40% of the paths into cases of type II and III, while less than 10% of cases exhibited a positive tail. Interestingly, we note that the percentage of samples falling into the Gaussian component for the full set of measured paths is over 85%. That is, most of the time the bandwidth time series follows a Gaussian distribution, oscillating softly over the mean, then, as an unusual event, bandwidth dips during a time (or even more infrequent, it peaks). We will refer to the Gaussian component as the stationary behavior of the phenomenon, which represents what a transfer can expect from a path in the Cloud. On the other hand, the excursions from the mean represent changes on the regular behavior. Especially of interest is when a time series dips, to which we often refer as to a downtime.

To formally split data into these two components, we simply apply a goodness-of-fit test on normality for the mean Gaussian component. In particular, we entrust Lilliefor's test at the 5% significance level with this task. Those samples that do not pass the test are considered as excursions.

### 2.4 Measurement campaign

So far, we have devised a testbed of eighteen data-centers that grouped in pairs makes up 306 paths to measure. The parameters for *cloudB* are samples of 5-minute duration, and we have fixed the sampling rate at one sample per hour to trade-off cost and thoroughness [22]. Finally, the measurement campaign took place during three weeks. By elaborating on the resulting set of data, we describe the main Gaussian component of an extensive set of bandwidth time series in the Cloud, and then, we relate such component to data-centers' locations, CSPs and peculiarities of specific data-centers. Moreover, we assess how downtimes (and peaks) of bandwidth time series are correlated. The two following sections focus on these two issues, respectively.

## 3 Stationary behavior

This section focuses on the stationary behavior of the bandwidth in the Cloud. First, we pay attention to the mean of its principal Gaussian component. Then, we study the factors that explain this behavior. In other words, we first describe the performance and then generalize the results.

### 3.1 Overview

Figure 6 shows the TCP bandwidth mean for the 306 paths under study, as well as the mean by data-center source (last column) and destination (last row), and overall mean in the Cloud (right and bottom -most square). Several observations arise. In the big picture, the overall mean in the Cloud ranges between 150 and 300 Mb/s, specifically, roughly 250 Mb/s. Although, the heterogeneity between different paths is clearly significant. On the one hand, some paths achieved rates close to 1 Gb/s although none of the them achieved a mean over 750 Mb/s. Actually, no single measurement exceeded 930 Mb/s, what supports that the capacity of the VMs on the testbed was adequate. On the other hand, several paths did not exceed the rate of 150 Mb/s. After a more detailed inspection, we assessed that all of the means surpassed 50 Mb/s.

We found some homogeneity by inspecting data-centers as sources. Three data-centers of $C1$ exhibit the best performance with averages of roughly 400 Mb/s, whereas the data-center of $C3$ located in East Asia is below 100 Mb/s. Regarding destinations, we identify more heterogeneity. $C2$ and $C4$ data-centers stand out, and Brazilian and Australian data-centers of $C1$ as well as the data-center of $C3$ in East Asia obtained modest results. By paying attention to the figure from a distance, some clusters become apparent. To mention some of them, the good relationship between $C1$ and $C4$ (top right part of the figure), the good overall performance of $C2$ and $C3$ (columns 7-13), and the better performance of $C1$ outside its own infrastructure (rows 1-6, columns 7-18). To formally identify these interactions, next we apply a factor analysis over the data.

As a precaution against possible congestion on VM interfaces, those paths with samples over 800 Mb/s were re-measured with enhanced VMs. Specifically, for an additional week after the measurement campaign was carried out. We did not find significant differences, and this set of measurements was not used in what follows.

### 3.2 Factor analysis of bandwidth time series

We have entrusted Analysis of Variance (ANOVA) with the factor analysis of the data. ANOVA is a widely used statistical methodology whereby the observed variance of a given response variable is split into explanatory factors or categories and provides a means to determine if the factors have any importance in explaining such a response variable, and how much this accounts for. For further details, see for instance [9, 19].

ANOVA methodology requires the data to meet a few requirements: First, the samples must be independent; second data must be Gaussian distributed; and third, they must fairly share the same intra-group variance (exhibit homoscedasticity). However, the results of ANOVA is generally accepted provided that the number of elements in each group is large and similar between them as well as there is not a far deviation from the homoscedasticity assumption in terms of mean-variance rate [26, 15, 14].

The data under study meets the first requirement as we are characterizing the main Gaussian component of the bandwidth time series as described in the previous section. A simple auto-correlation test proves that data is independent. Regarding, homoscedasticity is clear from the previous section that the wide of the Gaussian component differs across paths, a test as the Levene's one confirms this. However, note that the number of samples is large, balanced and CV is low as assumptions require.

Finally, ANOVA performs a contrast test using the ratio between the adjusted sum of squares of each factor and interactions, and the total, which follows a Snedecor-$F$ distribution under the null hypothesis; which considers that the total adjusted mean square is due to the experimental error, and not to differences in the population when grouped by factors. However, if the null hypothesis is not accepted, this means that the factor used to build the groups is statistically significant according to the $F$-test. In addition to the statistically significance, ANOVA also quantifies how much variance a factor explains. Specifically, the ratio between the sum of squares of samples that belong to each factor level, intra-level samples, and the total, inter-level samples. This is typically named explained variance by a factor or interactions of them.

In this regard by comparing the percentage of variation in the response variable that can be explained by the factors and interactions to the error, we will have a notion of the goodness of the model. That is, the coefficient of determination $\bar{R}^2$, which, typically, is considered as significant over 0.85.

### 3.3 Applying ANOVA to the data

We propose to follow an iterative approach, whereby the most general factors are first used to explain variance, and the most specific ones are progressively considered (ANOVA Type I). The target is to identify commonalities in the Cloud letting that that the most particular factors participate only for variance otherwise unexplained.

Formally, in this study, general factors are the geographical location or area (factor, $Area$) and the CSPs ($CSP$) as Table 1 details, and also the interaction between a pair of well-connected areas (factor $Area^S * Area^D$) or CSPs (factor $CSP^S * CSP^D$). The interactions between factors allow us to evaluate if the performance depends on the proximity of source and destination ($Area$), the connectivity is better inside CSPs' infrastructures ($CSP$) or, otherwise, each data-center behaves somewhat independently. On the other hand, the most particular factors or interactions are those related to data-center level. That is, factors that explain a time series according to the specific source data-center (factor $DC^S$) or destination data-center (factor $DC^D$), as well as a factor that accounts for both ends (factor $DC^S * DC^D$).

In addition to these factors, we also add two intrinsic ones [14]. Specifically, the time (UTC) the sample was gathered (factor, $Time$) and the day of the week (factor, $Weekday$). Actually, we consider them as the most general factors as they apply to all the measurements. Note that they allow us to investigate whether the capacity in the Cloud is better on certain times and days of the week.

By ordering factors from more general to more specific, we construct the following ANOVA model:

$$\begin{aligned} BW_{twijki'j'k'p} = C & \\ & + Time_t + Weekday_w \\ & + Area_i^S + CSP_j^S + DC_k^S \\ & + Area_{i'}^D + CSP_{j'}^D + DC_{k'}^D \\ & + Area_i^S * Area_{i'}^D \\ & + CSP_j^S * CSP_{j'}^D \\ & + DC_k^S * DC_{k'}^D \\ & + \epsilon_{twijki'j'k'p} \end{aligned} \quad (1)$$

where, $BW_{\bullet p}$ represents the $p^{th}$ observation (a bandwidth sample) that results of the addition of terms according to the $t^{th}$ level of factor $Time$ (e.g., 0 a.m., ... 12 p.m.), $w^{th}$ level of factor $Weekday$ (e.g., Mon, ... Sun), $i^{th}$ level of factor $Area$ as source (e.g., data-center source is located in East US, West US, ..., East Asia, see Table 1), $j^{th}$ level of factor $CSP$ being the source (e.g., $C1$, $C2$, $C3$, and $C4$), $k^{th}$ level of factor $DC$ being the source (e.g., Virginia$_{C1}$, California$_{C1}$, ... Taiwan$_{C4}$), and so forth for the levels for factors as destinations ($i'j'k'$). In addition, the two-way factors (represented by $*$) account for the impact that interactions exert. For example, the explained variance because of the source is East US and the destination is West US (including levels where both source and destination are the same). In more detail, an additional term for the interaction of $i^{th}$ and $i'^{th}$ levels of factor $Area$ (source and destination, respectively), for the interaction of $j^{th}$ and $j'^{th}$ levels of factor $CSP$ (again, source and destination) and finally another for the interaction of $k^{th}$ and $k'^{th}$ levels of factor $DC$ (also, source and destination).

In other words, $BW$ represents each of the samples ($p$), samples that are indexed using $t$ and $w$ by time and day of the week, as well as $i, j, k$ and $i', j', k'$ to index the geographical area, CSP, and data-center, both source ($S$) and destination ($D$) of the path, respectively. The intercept term or $\mu$ represents the overall mean response, that is, a constant figure over which the rest of factors add terms. Finally, the difference between a sample $p$ and the addition of the factor terms according to the levels of such sample (often named as model value), is typified by $\epsilon_{ijki'j'k'p}$ (often named as random or experimental error). In our case, $\epsilon$ is linked to the variance of the Gaussian component as Fig. 5 illustrated.

### 3.4 Results and discussion

Table 2 shows the results after applying the ANOVA test using SPSS [25] software over bandwidth time series measurements. First of all, the $R^2$ term is close to 1, so we can conclude that model explains the response variable (bandwidth in our case) with high accuracy. In this line, the last column exhibits the p-value for the null hypothesis that supports the homogeneity of means, such hypothesis can be reject with significant confidence, i.e., all factors are able to explain some variance.

However not all of them explain variance in a similar amount. The second column (sum of squares in ANOVA terminology) represents the explained variance in absolute terms, while the third column shows these figures as percentages. This latter can be interpreted as the percentages of variance that each factor or interaction of them help to explain. To make the data easier to contrast, the fourth column shows the percentage that each row accounts once the intercept term is not considered. Note that percentage that the intercept term represents a significant fraction of the bandwidth of each path (the next section will estimate this term in roughly 40 Mb/s).

After that, we find that factors such as sources, $Area^S$ and $CSP^S$, depict modest significance in comparison to their destination counterparts $Area^D$ and $CSP^D$ that account for larger figures. This implies that the bandwidth is especially sensitive to where the path is destined. Even more, the two-way factors involving pairs of $Area$ and $CSP$ levels account for more than 30% of the explained variance apart from the intercept term.

So far, before even considering the data-center ends, almost 88% of the total variance is explained. This highlights that the bandwidth behavior strongly depends on data-centers location and CSP evolved, rather than a specific behavior of data-centers themselves. We speculate that this is due to the diverse routing agreements that govern the Internet, and to the well-known relationship latency/TCP bandwidth. The closer two nodes are, the lower latency is, and so that higher the TCP bandwidth. The data-center factors itself explain less than %5 of variance, and only when both source and destination data-centers are considered this figure increases to roughly 6% of the total variance.

Finally, the factor $Time$ only exhibited marginal significance, and no qualitative importance (less than 0.0%). This implies that the Cloud is almost insensitive to time, reasonable in the Cloud as it encompasses users and nodes widespread in the world. The $Weekday$ factor shows modest qualitative importance (%0.4), although as will be shown, according to the parameter estimates the difference between weekends and working days is clear. This contrasts with other commercial and academic networks with a both strong daily and weekly patterns.

Given these results, the idea of extrapolate the bandwidth for non-measured paths based on factors makes sense. This is useful both for estimating performance of un-deployed nodes and avoiding gathering measurements for each of the paths of a large deployment. Resulting in potential savings by reducing monitoring systems' costs.

### 3.4.1 Inspecting the levels

While the ANOVA table has unraveled how factors influence bandwidth time series, the parameter estimates will shed light on how their different levels interacts. That is, ANOVA identified that factors $Area$ and $CSP$ can explain much of the bandwidth time series, but what specific values such factors take to increase or decrease the bandwidth is a matter of the parameter estimates. Intuitively, performance should peak when a path encompasses the same area and same CSP. Let see to what extent this is true.

Table 3 summarizes the parameter estimates for the set of general factor $\mu$, $Time$, $Weekday$, $Area$, $CSP$ and these two latter factors interactions. The parameters estimates of such general factors allow us not only to discuss the interaction between levels, but their explicit inclusion provides the Internet community with an extrapolate set of data in the Cloud to consider in their research and commercial tasks. We do not include data-centers factors as there are 306 possible combinations, and because the addition of all the terms of the model simply equals to the values depicted in Fig. 6.

The parameters for factor *Time* levels unveil that Cloud is practically insensitive to the time throughout the day, with a minimum at 3 p.m. (UTC) and peaks at 2 a.m. and 11 p.m. It is difficult to relate this to properties of human activity as multiple Time-zones are involved, indeed likely this is because the influence is low. By turning the focus on the day of the week, the homogeneity between two groups, working days and weekends becomes apparent. The largest values (i.e., more capacity and intuitively less use of the Cloud infrastructure) are those from weekends with an additional term of about 30 Mb/s. It is worth remarking that Mondays behave similarly to weekends likely because they include a fraction of the previous Sunday in some parts of the world at UTC time.

The *Area* factors as source and destination represent an additional term of 85 Mb/s in some cases, among which the data-centers located in both coasts of US stand out. On the other hand, the data-centers in South America exhibit the lowest parameters. Regarding CSPs, $C4$ data-centers share the largest capacities in both directions of traffic, and $C1$ and $C3$ show more moderate rates.

Finally, some interesting conclusions arise out of the two-way factors. Specially, when we pay attention to those rows where levels are equal (e.g., $C1 * C1$). They represent the additional terms for paths whose ends are either in the same CSP or the same Area. The pairs with the same CSP do not show larger values, even for $C1$*$C1$ and $C4$*$C4$ pairs the addition that such combinations represent is zero (i.e., no additional bandwidth because of such combination). On the contrary, paths from $C1$ to $C4$ have an increment of more than 300 Mb/s, although the increment in the inverse direction is much lower. In summary, we conclude that the pair of $CSP$ involved in a transfer is important, but somewhat counter-intuitively, not because data-centers belong to the same CSP. We speculate that the reasons for this can be the different agreements for Internet routing each CSP has, i.e., apparently not many bilateral agreements exist.

The impact of the two-way *Area* factor is just the opposite. The terms to add when two data-centers are in the same area are large figures, specifically they range from 105 to 433 Mb/s (East Asia and South America, respectively). In between, the rest of the areas connect internally with an addition term of about 300 Mb/s. By inspecting the cross relationships, it becomes apparent that East Asia is in some way isolated. Terms that represent paths to/from East Asia hardly have any additional bandwidth, while the other cross terms (connections between areas apart from East Asia) are above the range of 100 Mb/s. In conclusion, the bandwidth-delay product seems to play an import role in the Cloud, as the proximity between data-center boost performance.

Finally, regarding data-center factors, we note that in all cases the additional term that represent such factors is lower than 100 Mb/s, and typically closer to zero than such 100 Mb/s rate. After ordering these factors by impact, pairs of data-centers that stand out are those in Ireland and pairs of data-centers located one in Central US and another in one of the US coasts. In these cases, a term only slightly below 100 Mb/s is added.

# 4 Correlation across paths

We turn our attention to the correlation between the full set of measured paths. In other words, while the previous section focused on the principal Gaussian component in a stationary viewpoint, we now focus especially on the excursions. To this end, we have calculated the Pearson correla-tion coefficient ($\rho$) [8] between all the bandwidth time series (respecting the synchronism between signals).

Should excursions occur at same time, $\rho$ will be significant, otherwise, the coefficient tends to be close to zero so indicating that bandwidth change independently between paths. We note that a positive $\rho$ implies that two time series change in the same way. That is, when a path performance peaks the other does the same, when one works poorly the other too. On the other hand, non-significant coefficients and, especially, negatives ones will point at paths that can provide availability and reliability to others in the Cloud. Essentially, a downtime in a path can be resolved by the good performance, or at least a regular one, of an alternative path.

In this light, we pose two questions: (i) if the performance between data-centers in the same area is correlated, which would support the approach of achieving redundancy by disseminating data in different areas; (ii) if data-centers in the same CSP are correlated, which would entail additional advantages (apart from lower latencies [33]) of multi-CSP deployments. To answer them, we follow an equivalent approach to that in the previous section.

## 4.1 Data analysis

Our testbed is comprised of 18 data-centers with, consequently, 306 paths between them which in turn translates into 93,330 possible coefficients, one per each possible pair of paths in such a set of 306 paths.

To provide an overview, Fig. 7 shows the $\rho$ as a ECDF for the full set of coefficients. Considering that values between -0.25 and 0.25 as no correlated, the figure shows that about 80% of the pair of paths are not so. Then, roughly a 10% of the tests yielded clearly positive in correlation, let say figures lower than -0.5 and higher than 0.5. As an attempt to identify those paths, Fig. 8 shows the mean of $\rho$ per each path indexed by source and destination data-centers. For example, given the path from Virginia$_{C1}$ to California$_{C1}$, the cell (1,2) represents the average $\rho$ of this path to the rest of the paths. In turn, the last column and row average $\rho$ by source and destination data-center, respectively. In the figure, the coefficients are separated into five classes that we relate to strong correlation, significant correlation (both in negative and positive directions), and marginal/no correlation. In general, these initial results suggest that the correlation is low.

By paying attention to the last column, in average it becomes apparent that none of the data-centers stands out as particularly sensitive to other paths changes. That is, in the big picture there is no evidence of that bandwidth changes in unison in the Cloud. However, the same figure shows that some paths had correlation, and importantly, the effect of averaging may have hidden details. A factor analysis will highlight what has in common the pair of correlated paths. We note that a full-factorial approach to this problem would include 4-way factors making the model hardly tractable (i.e., interaction of four factors, specificly, both source and destination of the path under study, and also both source and destination of the path to be compared). However, our focus is to shed light on if a data-center may expect correlation between the set of paths it can potentially use. That is, if a path in use is working poorly, another path can compensate for that, and this is only possible if the source of the paths are the same node. In other words, to study the correlation between paths in which the source is the same node, in contrast to compare all paths including disjoint ones.

We then simplify the problem, resulting in the following model:

$$
\begin{aligned}
CORR_{ijki'j'k'i''j''k''p} = C \\
&+ Area_i^S + CSP_j^S + DC_k^S \\
&+ Area_{i'}^{D1} + CSP_{j'}^{D1} + DC_{k'}^{D1} \\
&+ Area_{i''}^{D2} + CSP_{j''}^{D2} + DC_{k''}^{D2} \\
&+ Area_i^S * Area_{i'}^{D1} \\
&+ Area_i^S * Area_{i''}^{D2} \\
&+ Area_{i'}^{D1} * Area_{i''}^{D2} \\
&+ CSP_j^S * CSP_{j'}^{D1} \\
&+ CSP_j^S * CSP_{j''}^{D2} \\
&+ CSP_{j'}^{D1} * CSP_{j''}^{D2} \\
&+ DC_k^S * DC_{k'}^{D1} \\
&+ DC_k^S * DC_{k''}^{D2} \\
&+ DC_{k'}^{D1} * DC_{k''}^{D2} \\
&+ Area_i^S * Area_{i'}^{D1} * Area_{i''}^{D2} \\
&+ CSP_j^S * CSP_{j'}^{D1} * CSP_{j''}^{D2} \\
&+ DC_k^S * DC_{k'}^{D1} * DC_{k''}^{D2} \\
&+ \epsilon_{ijki'j'k'i''j''k''p} \qquad (2)
\end{aligned}
$$

where the same terminology that Equation 1 is followed, apart from that we are modeling correlation coefficients ($CORR$) and that we are now considering three data-center ends. A source data-center ($S$) and the two destination data-centers ($D1$ and $D2$), then the correlation coefficient is calculated between the paths $S$-$D1$ and $S$-$D2$, for all possible combinations of data-centers (or levels), where $'$ and $''$ will be used to index levels within the two destination factors, respectively. Finally, the 2-way factors account for the correlation to add because of interactions of data-centers levels taken in pairs, and the 3-way factors for an additional terms because of the combinations of levels as trios of data-centers.

## 4.2 Results and discussion

Many observations arise by inspection the percentage of variance that each factor explains, Table 4 (third column). First, the *Area* and *CSP* as destinations turn out to be no significant, while these factors understood as sources are clearly significant. That implies that certain areas and CSPs are less sensitive to changes in the Cloud (i.e., more robust).

However, the explained variance of 2-way and 3-way factors that involve *Area* and *CSP* factors, yet some of them statistically significant, is qualitatively marginal. In practical terms, we have found no evidence of that data-centers within the same area or CSP present different correlation between them that with respect to other data-center in the Cloud. In this way, the search for availability and reliability (note that latency or other QoS metrics are a different matter) in the Cloud can be achieved in the same manner by close and far-away data-centers, and only moderately improved by spreading data between different CSPs rather than relying on the same CSP. On the other hand, the factors and interactions related to data-centers are qualitatively very significant. In conclusion, each pair of paths is correlated in a data-center basis with little contribution from the pair of locations and CSPs involved. The impact of this on current or future deployments is in terms of making the reliability simpler. But, unfortunately, it also makes more complex to infer the interactions between path based on general factors as we did with the bandwidth, as in this case, pairs of similar paths behave heterogeneously. Although, in general the correlation between paths is relatively low in the Cloud as the figures 7 and 8 proved.

The parameter estimates for factors $Area^S$ and $CSP^S$ are shown in Table 5 as the only general factors with some qualitative importance. We found that both US costs present the smallest measured $rho$ which translates into greater robustness to overall changes in the Cloud. Also Australia and North Europe showed low estimations for $rho$, while East Asia and Central US areas exhibit an additional term of more than 0.3. That is, these latter areas comprise more sensitive data-centers to oscillation on the bandwidth capacity in the Cloud in its whole. Similarly, while CSP $C4$ presented an estimated additional correlation of zero when transfers are sourced in one of its data-centers, $C3$ showed the largest positive $rho$ figure. After inspecting the parameters estimates for the full data-center factors and its interactions, the lower $rho$ of data-centers within Ireland, and the high estimates for $\text{Europe}_{C4}$ stand out. We also remark the low correlation between data-centers located in Virginia, and a generally high correlation of the data-centers located in Singapore.

# 5 Related work

In this section, we first review studies that measured bandwidth of infrastructures comparable to the public Cloud. Then, studies focused on measuring the Cloud from different points of view (among others metrics, VM flavors, storage capacities, latency, popularity, and CPU) or that reported some Cloud measurements with other final aims.

Regarding other infrastructures, the authors in [21] measured the bandwidth between most of the nodes of the Planetlab platform. Planetlab is a distributed federation where members can take the control of probes across the world to test any novel algorithm or idea. In this case, the authors leveraged such platform to measure the bandwidth between parts of the world, specifically 250 probes were deployed and its connectivity tested. They entrusted Pathrate with this task, tool that estimates the maximum capacity of a path and not the available TCP bandwidth as we devised. Unfortunately, they found a set of stricter bandwidth limits on probes that polluted the measurement campaign. That is, some of the measurements are simply bounded to such limits without any link to the real capacity between end nodes. This justifies the careful selection of VM probes in our testbed. Apart from the capped samples, they found that paths typically range between 80 and 120 Mb/s, clearly below our measurements in the Cloud.

With the final target of finding overlay routes in the Cloud to improve quality of multimedia services, the authors in [10] showed some figures about TCP throughput measurements in the Cloud. Their solution was to disseminate contents but, importantly, with the constraint of not increasing costs. To do that, they limited the throughput of the processes to distribute data to the original $95^{th}$ percentile of the network (typically, the metric used to charge CSP). To test their ideas, they measured the bandwidth between data-centers in Amazon's infrastructure. Specifically, between data-centers in North California, Oregon, Virginia, Sao Paulo, Ireland, Singapore and Tokyo. This set differs from ours in a couple of data-centers in Amazon apart from all the data-centers of other CSPs we study. They measured bandwidths for 3 minutes per each path, but they neither specify which tool was used nor any other details, so limiting the comparison. In general, they found lower bandwidths than ours.

Paying attention to the careful selection of probe capacities, the authors in [31] studied the impact of the virtualization on TCP throughput at the data-centers of Amazon located in East US. They found that virtualization exerts an impact when several VMs share the same CPU as periodically the sender is taken out of the CPU and throughput decreases. To achieve such a conclusion, they compared at low-scale the TCP throughput of both small and medium instances. After this experiment, they found that small instances achieved up to 500 Mb/s of TCP throughput whereas the medium probes achieves almost 900 Mb/s.

Similary, the authors in [16] studied how to systematically pick a combination of CPU and VMs flavors to match the requirements of applications in the Cloud. They consider that bandwidth is one of the possible requirements a application must meet, and measured the bandwidth between different VM flavors at Amazon EC2 both inside and outside the same data-center. Their results conclude that the medium and large VMs achieved bandwidths close to 1 Gb/s when both ends are in the same data-center. On the other hand, wide-area bandwidths exhibited figures far lower bandwidths, typically bounded by 200 Mb/s. We have found larger values for EC2 but in any case clearly below the capacity of the VM interface, in general below some 300 Mb/s.

Amazon's cloud infrastructure was also studied in [30], paper that spans metrics such as latency and CPU performance using benchmarks. Differently, in [3], such Amazon infrastructure was examined from the perspective of end-users. Specifically, they found that many current deployments on Amazon are not yet exploiting the geo-distribution of contents for better quality of service. As an example, users from Italy are mostly served by the data center located in Virginia instead of its counterpart in Ireland. Even more, when the perceived quality for end-users in Italy was considered poor.

Finally, an extensive effort to characterize the Cloud was presented in [22]. The authors presented the results of applying the monitoring tool CloudCmp in several CSPs as of 2010. They paid much attention to metrics such as storage performance, CPU capacity, latency between different CSPs and intra-data-center networking. However, they did not study the inter-data-center bandwidth on an extensive set of nodes across different areas. Specifically, they paid attention to data-centers located at US, and reported that the bandwidth between them ranges between 100 Mb/s and 500 Mb/s. Our measurements are in line with theirs.

As distinguishing characteristics, we note that we have extended our study to diverse set of data-centers from several geographical areas and CSPs in attempt to provide general results for the public Cloud. It is also worth remarking that we measured the bandwidth in terms of available TCP throughput as an approximation to the quality perceived by transfers. Moreover, we have shed light on how to measure in the Cloud to provide both a significant description and modeling of the measurements. Indeed, by carrying out a systematically measurement campaign, we have been able to give a novel comparison between data-centers' behaviors based on factors (especially, general ones). Such factors have revealed interesting conclusions otherwise hidden in the data. Finally, to the best of our knowledge, the study of the correlation of bandwidth time series has not been previously addressed, despite its clear link to availability and reliability in the Cloud.

# 6 Conclusions and future work

Throughout this paper, we have shed light on the important metric of bandwidth within the public Cloud. Besides a comparison of the Cloud performance of interest for CSPs and a further step on the characterization of the Internet of interest for the research community, we believe we have provided Cloud costumers with formal indications and generalities of what they can expect when openning a TCP connection on their deployments. And importantly, both in terms of regular and malfunction behaviors.

That is, we have proposed a two-component approach to model time series of bandwidth on the Cloud. These components account for, on the one hand, the stationary behavior of data-centers (i.e., how data-centers tend to work), and, on the other hand, unexpected excursions (typically downtimes, although also peaks).

While the importance of the stationary behavior is immediate, the study of downtimes is no less transcendent. The question of whether changes on the bandwidth capacity across the Cloud tend to be a synchronized process, or conversely, data-centers operates at its own, has a dramatic impact on the robustness of any deployment in the Cloud.

After applying a factor analysis on an extensive testbed, we have concluded that the behavior of the stationary component can be considered as a homogeneous phenomenon, whereas the same does not apply to the other component. That is, while according to the locations and the CSPs involved as source and destination in a transfer most of the bandwidth time series' variance can be described, the correlation between such time series (where the excursions are especially of importance) is a process that qualitatively depends on the specific data-center ends involved.

By examining the specific parameters that increase the bandwidth of paths, we have found that transfers inside the same geographical area receive a significant additional term, while the same does not apply inside CSPs. In more detail, the source CSP is significant in explaining the bandwidth of a path, but that a path involves the same CSP as source and destination is only of marginal importance. As other interesting conclusions, we have found that time in which the Cloud is measured is of little significance, and the week day only in a modest manner. That is, the behavior differs from weekends to working days but not between them. This has an impact on how to schedule bulk transfers in the Cloud.

The correlation study also exerts a direct impact on the planning of Cloud deployments. Fortunately, the low correlation found and the moderate significance of CSP and location data-center interaction factors eases the achievement of robustness, as close data-centers and data-centers inside same CSP did not show more correlation that those located far away belonging to other CSPs. However, the peculiarities of data-centers play an important role so that the generalization of the results is more limited. Consequently, continuous fine-grain monitoring at multiple probes is needed to estimate this metrics.

We believe that all these lessons learned and the measurements reported along the paper are of interest for both practitioners and researchers. The measurements and their description are useful to make decisions for current and future deployments, and even to extrapolate figures to non-measured data-centers pairs with certain confidence. Similarly, we see the conclusions and the identification of invariants as a step toward better knowledge of the dynamics of the Cloud, and, consequently of the Internet.

As future work, we plan to carry out a formal evaluation of error that the extrapolation of measurements to non-monitored data-centers entail; as well as to include the impact that the potential heterogenous performance that even equal-equipped VMs with the same CSPs can exert on the generalization of measurements in the Cloud.

## Acknowledgements

This research was carried out with the support of the Prometeo project of the Secretariat for Higher Education, Science, Technology and Innovation of the Republic of Ecuador. It was also partially supported by the UTN-CUICYT-177 project Characterization of Correlated Performance in the Cloud. Finally, the author thanks professors Edgar A. Maya and Julio J. Armas for their assistance.

## References

[1] Alexa. Top sites in the cloud. `http://www.alexa.com/topsites/category/Top/Computers/Internet/Cloud_Computing` [1 Dic 2015], 2015.

[2] David Andersen, Hari Balakrishnan, Frans Kaashoek, and Robert Morris. Resilient overlay networks. *SIGOPS - Operating Systems Review*, 35(5):131–145, 2001.

[3] I. Bermudez, S. Traverso, M. Mellia, and M. Munafò. Exploring the cloud from passive measurements: The Amazon AWS case. In *Proceedings IEEE INFOCOM*, pages 230–234, 2013.

[4] A. W. Bowman and A. Azzalini. *Applied Smoothing Techniques for Data Analysis.* Oxford University Press Inc., 1997.

[5] Rodrigo N. Calheiros, Rajiv Ranjan, Anton Beloglazov, C. A. F. De Rose, and Rajkumar Buyya. Cloudsim: A toolkit for modeling and simulation of cloud computing environments and evaluation of resource provisioning algorithms. *Software: Practice and Experience*, 41(1):23–50, 2011.

[6] k. claffy. *Internet traffic characterization.* PhD thesis, UC San Diego, Jun 1994.

[7] James C. Corbett, Jeffrey Dean, Michael Epstein, Andrew Fikes, Christopher Frost, and et al. Spanner: Google's globally-distributed database. In *USENIX conference on Operating Systems Design and Implementation*, pages 251–264, 2012.

[8] Mark Crovella and Balachander Krishnamurthy. *Internet measurement: infrastructure, traffic and applications.* John Wiley and Sons, Inc., 2006.

[9] Olive Jean Dunn and Virginia A. Clark. *Applied Satistics: Analysis of Variance and Regression.* John Wiley and Sons Inc., 1974.

[10] Yuan Feng, Baochun Li, and Bo Li. Jetway: minimizing costs on inter-datacenter video traffic. In *ACM international conference on Multimedia*, pages 259–268, 2012.

[11] Sally Floyd and Vern Paxson. Difficulties in simulating the Internet. *IEEE/ACM Transaction on Networking*, 9(4):392–403, 2001.

[12] Forrester Research. The future of data center wide-area networking. `http://www.forrester.com` [1 Dic 2015].

[13] J. L. Garcia-Dorado and S. G. Rao. Cost-aware multi data-center bulk transfers in the cloud from a customer-side perspective. *IEEE Transactions on Cloud Computing*, 2015.

[14] José Luis García-Dorado, José Alberto Hernández, Javier Aracil, Jorge E. López de Vergara, and Sergio Lopez-Buedo. Characterization of the busy-hour traffic of IP networks based on their intrinsic features. *Computer Networks*, 55(9):2111–2125, 2011.

[15] G. V. Glass, P. D. Peckham, and J. R. Sanders. Consequences of failure to meet assumptions underlying the fixed effects analysis of variance and covariance. *Review of Educational Research*, 42(3):237–288, 1972.

[16] Mohammad Hajjat, Ruiqi Liu, Yiyang Chang, T S Eugene Ng, and Sanjay Rao. Application-specific configuration selection in the cloud: impact of provider policy and potential of systematic testing. In *IEEE INFOCOM*, 2015.

[17] High Scalability. Latency is everywhere and it costs you sales - how to crush it. `http://highscalability.com/latency-everywhere-and-it-costs-you-sales-how-crush-it/` [1 Dic 2015].

[18] Manish Jain and Constantinos Dovrolis. End-to-end available bandwidth: Measurement methodology, dynamics, and relation with TCP throughput. In *ACM SIGCOMM*, pages 295–308, 2002.

[19] R. Jain. *The Art of Computer System Performance Analysis*. John Wiley and Sons Inc., 1991.

[20] Avinash Lakshman and Prashant Malik. Cassandra: a decentralized structured storage system. *SIGOPS - Operating Systems Review*, 44(2):35–40, 2010.

[21] Sung-Ju Lee, Puneet Sharma, Sujata Banerjee, Sujoy Basu, and Rodrigo Fonseca. Measuring bandwidth between Planetlab nodes. In *Passive and Active Network Measurement*, pages 292–305, 2005.

[22] Ang Li, Xiaowei Yang, Srikanth Kandula, and Ming Zhang. Cloudcmp: comparing public cloud providers. In *ACM SIGCOMM Conference on Internet measurement*, pages 1–14, 2010.

[23] Ajay Mahimkar, Angela Chiu, Robert Doverspike, Mark D. Feuer, Peter Magill, Emmanuil Mavrogiorgis, Jorge Pastor, Sheryl L. Woodward, and Jennifer Yates. Bandwidth on demand for inter-data center communication. In *ACM Workshop on Hot Topics in Networks*, pages 24:1–24:6, 2011.

[24] Thyaga Nandagopal and Krishna P. N. Puttaswamy. Lowering inter-datacenter bandwidth costs via bulk data scheduling. In *Symposium on Cluster, Cloud and Grid Computing*, pages 244–251, 2012.

[25] Norman H Nie, Dale H Bent, and C Hadlai Hull. *SPSS: Statistical package for the social sciences*, volume 227. McGraw-Hill New York, 1975.

[26] Stephen F. Olejnik and James Algina. Parametric ANCOVA and the rank transform ANCOVA when the data are conditionally non-normal and heteroscedastic. *Journal of Educational Statistics and Behavioral Statistics*, 9(2):129–149, 1984.

[27] P.N. Shankaranarayanan, A. Sivakumar, S. Rao, and M. Tawarmalani. Performance sensitive replication in geo-distributed cloud datastores. In *IEEE/IFIP International Conference on Dependable Systems and Networks*, pages 240–251, 2014.

[28] A. Tirumala, M. Gates, F. Qin, J. Dugan, and J. Ferguson. Iperf - the TCP/UDP bandwidth measurement tool. `http://sourceforge.net/projects/iperf/` [1 Dic 2015].

[29] Remco van de Meent, Michel Mandjes, and Aiko Pras. Gaussian traffic everywhere? In *IEEE ICC*, volume 2, pages 573–578, 2006.

[30] Edward Walker. Benchmarking amazon EC2 for high-performance scientific computing. *Login: the magazine of USENIX & SAGE*, 33(5):18–23, 2008.

[31] Guohui Wang and T S Eugene Ng. The impact of virtualization on network performance of Amazon EC2 data center. In *Proceedings IEEE INFOCOM*, pages 1–9, 2010.

[32] Timothy Wood, K. K. Ramakrishnan, Prashant Shenoy, and Jacobus van der Merwe. Cloudnet: dynamic pooling of cloud resources by live WAN migration of virtual machines. *ACM SIGPLAN Notices*, 46(7):121–132, March 2011.

[33] Zhe Wu and Harsha V. Madhyastha. Understanding the latency benefits of multi-cloud webservice deployments. *Computer Communication Review*, 43(2):13–20, April 2013.

[34] Eyal Zohar, Israel Cidon, and Osnat (Ossi) Mokryn. The power of prediction: cloud bandwidth and cost reduction. In *ACM SIGCOMM*, pages 86–97, 2011.

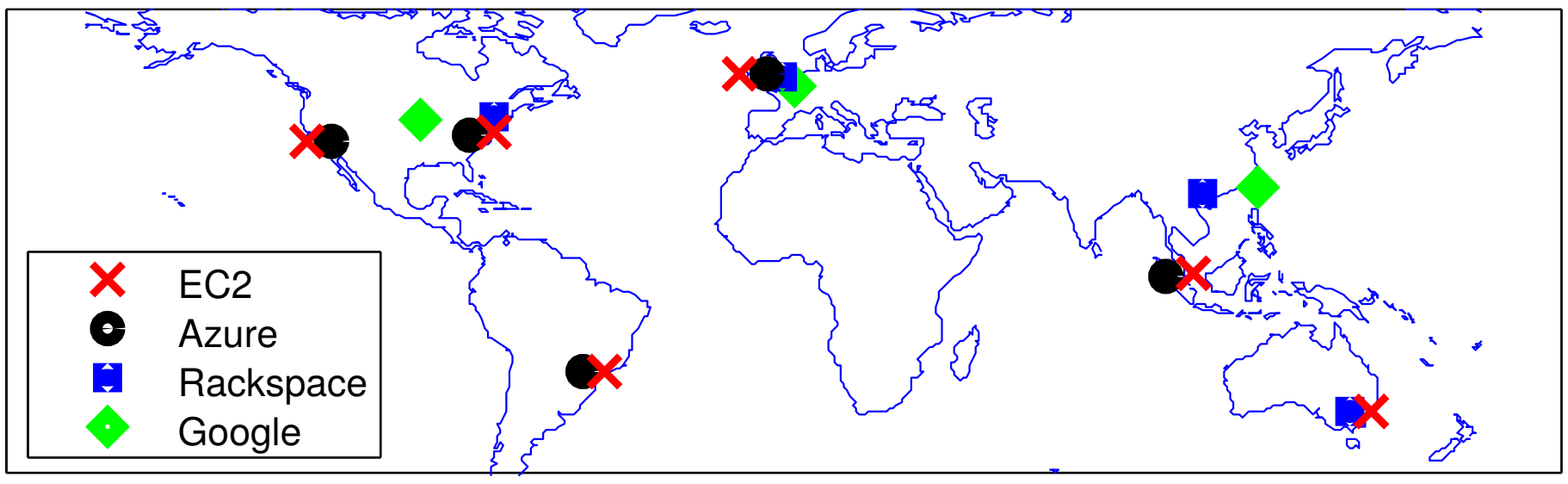

Figure 1: Multi-CSP and multi data-center testbed deployment

Table 1: Description of data-centers under study

| CSP | Abrv. | Geo. area | Data-center name |
|---|---|---|---|
| Amazon EC2 | $C1$ | East US | $\text{Virginia}_{C1}$ |
| | | West US | $\text{California}_{C1}$ |
| | | North Europe | $\text{Ireland}_{C1}$ |
| | | East Asia | $\text{Singapore}_{C1}$ |
| | | Australia | $\text{Sydney}_{C1}$ |
| | | South America | $\text{Sao Paulo}_{C1}$ |
| Microsoft AZURE | $C2$ | East US | $\text{Virginia}_{C2}$ |
| | | West US | $\text{California}_{C2}$ |
| | | South America | $\text{Brazil}_{C2}$ |
| | | North Europe | $\text{Dublin}_{C2}$ |
| | | East Asia | $\text{Singapore}_{C2}$ |
| Rackspace Cloud | $C3$ | East US | $\text{Virginia}_{C3}$ |
| | | North Europe | $\text{London}_{C3}$ |
| | | East Asia | $\text{Hong Kong}_{C3}$ |
| | | Australia | $\text{Sydney}_{C3}$ |
| Google Cloud | $C4$ | Central US | $\text{Central}_{C4}$ |
| | | North Europe | $\text{Europe}_{C4}$ |
| | | East Asia | $\text{Taiwan}_{C4}$ |

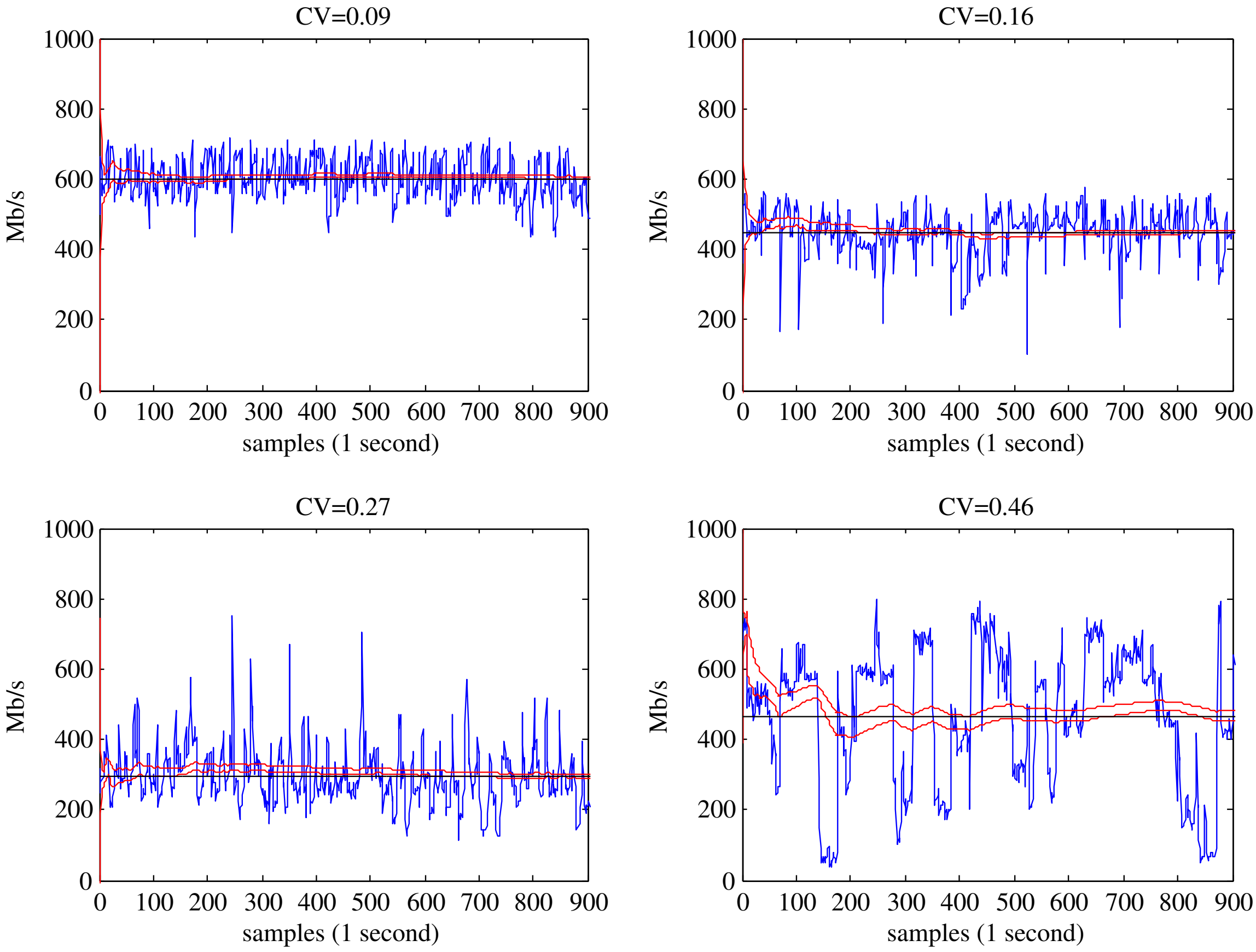

Figure 2: Examples of bandwidth time series behaviors according to its CV

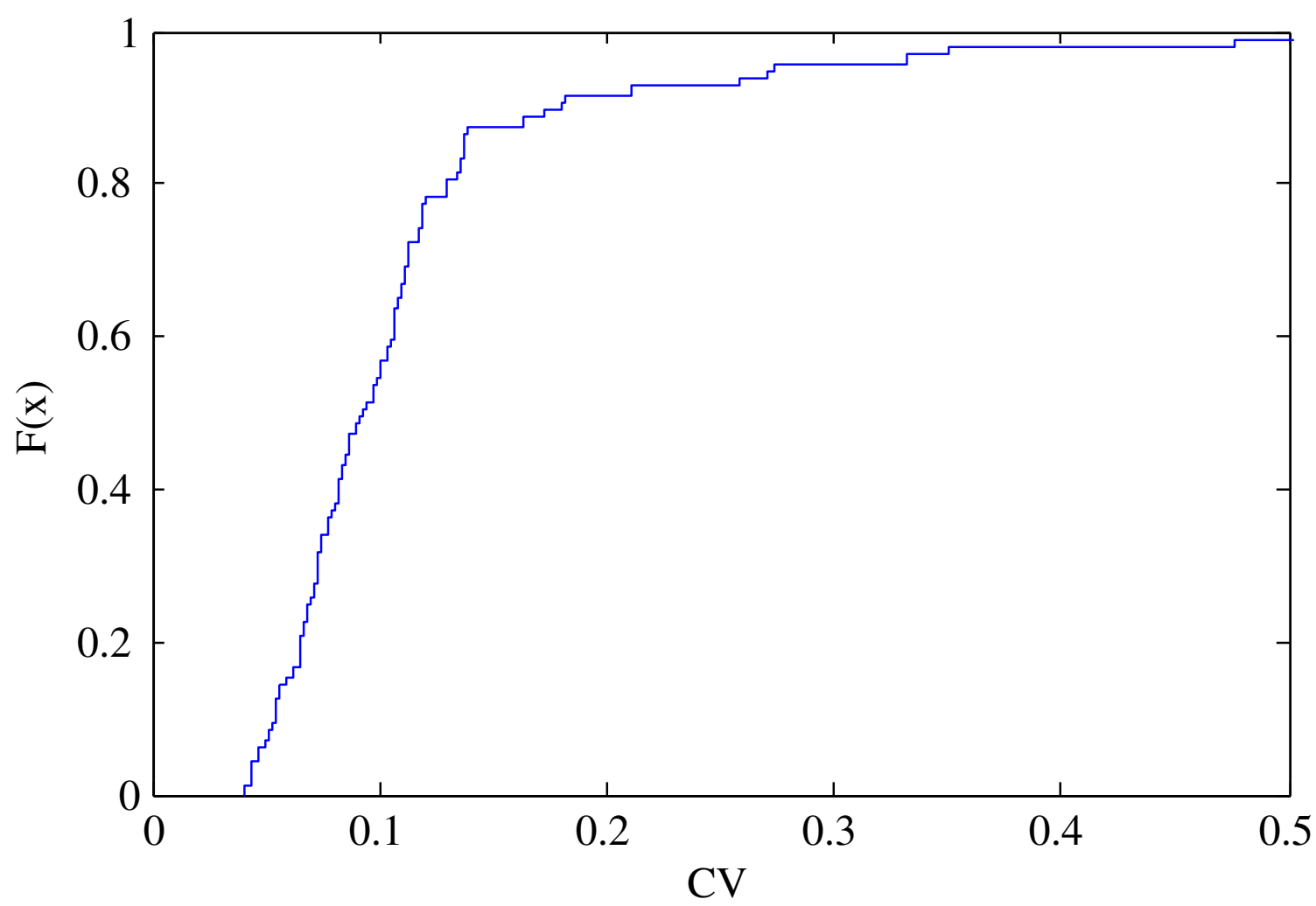

Figure 3: Empirical cumulative distribution function for the CV of the sample

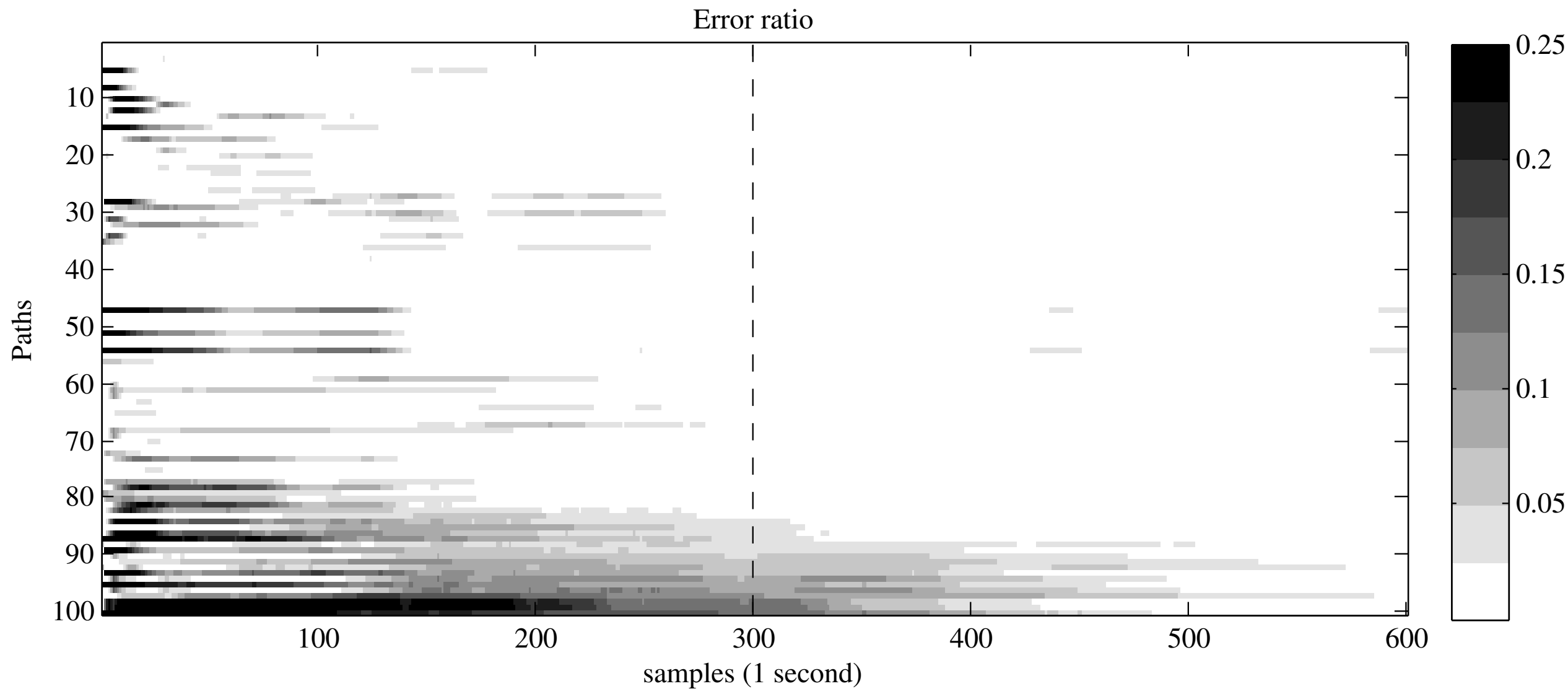

Figure 4: Error ratio (paths by ascending order)

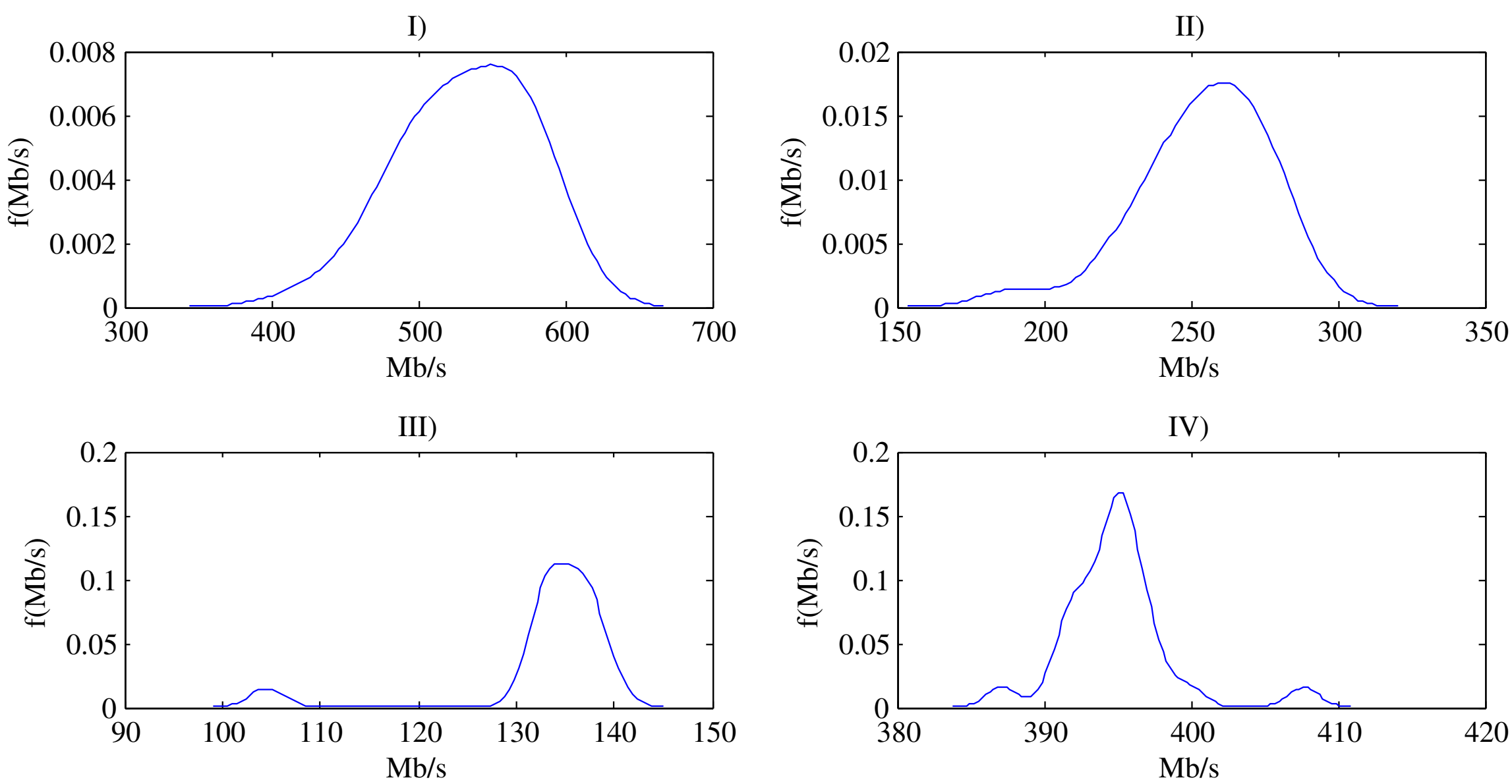

Figure 5: Examples of bandwidth time series behaviors

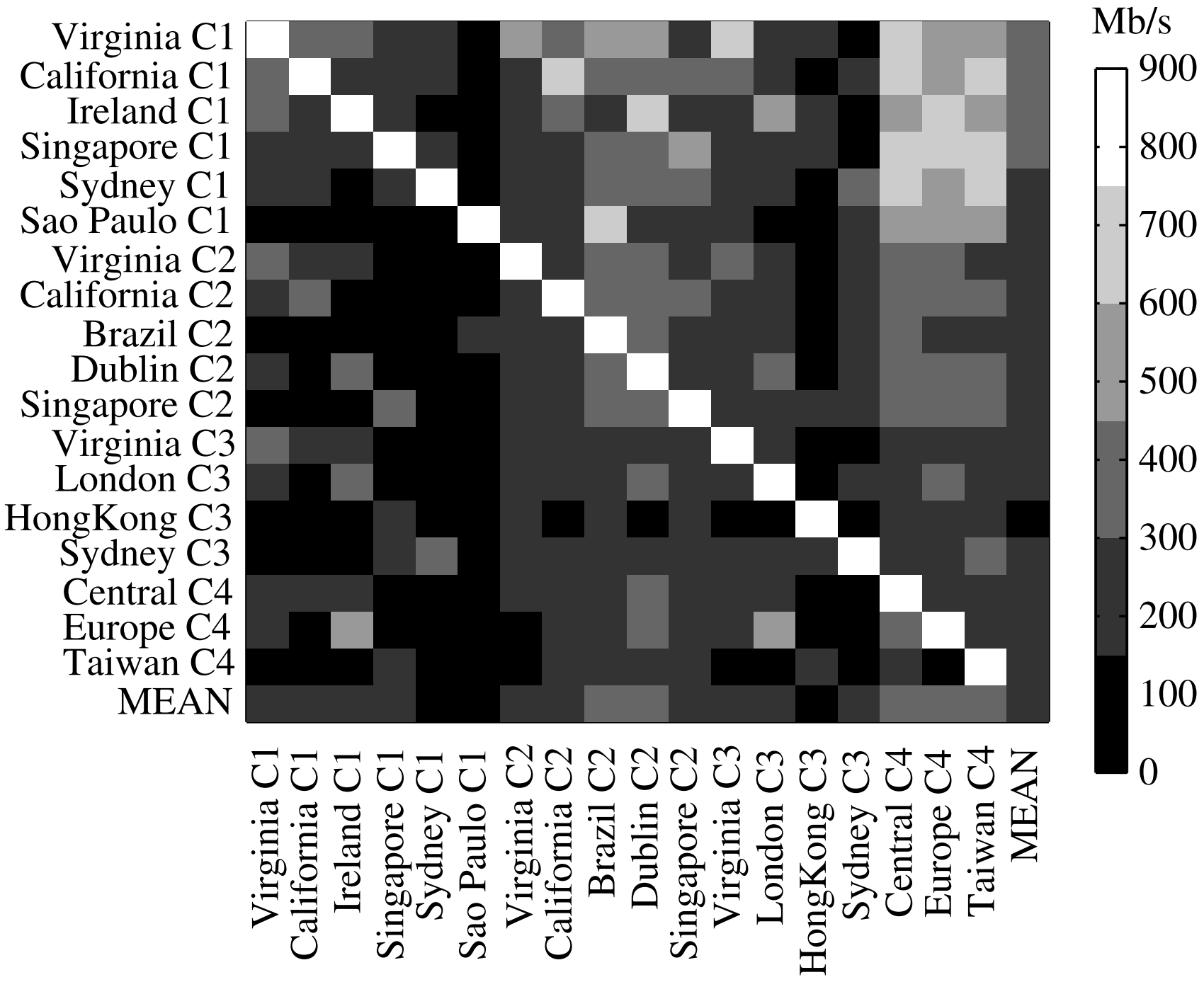

Figure 6: TCP bandwidth mean per path between source (vertical) and destination (horizontal) data-centers

Table 2: ANOVA table with $Time$, $Weekday$, $Area$, $CSP$ and $DC$ (data-center) and significant interactions as fixed factors, with bandwidth time series as response variable

Response variable: Bandwidth time series

| Factor | Sum of Squares | %Total | %Factors | df | Mean Square | F | $p$-value |
|---|---|---|---|---|---|---|---|
| $\mu$ | 1009168371 | 49.2 | · | 1 | 1009168371 | 169459 | 0.00 |
| $Time$ | 1105358 | 0.0 | 0.1 | 23 | 48059 | 8 | 0.01 |
| $WeekDay$ | 7991054 | 0.4 | 0.8 | 6 | 1331842 | 224 | 0.00 |
| $Area^S$ | 41652076 | 2.0 | 4.0 | 6 | 6942013 | 1166 | 0.00 |
| $Area^D$ | 125808840 | 6.1 | 12.1 | 6 | 20968140 | 3521 | 0.00 |
| $CSP^S$ | 80548490 | 3.9 | 7.7 | 3 | 26849497 | 4509 | 0.00 |
| $CSP^D$ | 176277823 | 8.6 | 16.9 | 3 | 58759274 | 9867 | 0.00 |
| $DC^S$ | 8274219 | 0.4 | 0.8 | 8 | 1034277 | 174 | 0.00 |
| $DC^D$ | 40486106 | 2.0 | 3.9 | 8 | 5060763 | 850 | 0.00 |
| $Area^S * Area^D$ | 192441930 | 9.4 | 18.5 | 35 | 5498341 | 923 | 0.00 |
| $CSP^S * CSP^D$ | 125439547 | 6.1 | 12.0 | 9 | 13937727 | 2340 | 0.00 |
| $DC^S * DC^D$ | 75897282 | 3.7 | 7.3 | 226 | 335829 | 56 | 0.00 |
| Error | 167088766 | 8.1 | 16.0 | 41491 | 4027 | | |
| Total | 2052179862 | 100 | 100 | 41825 | | | |

$R^2$=0.919

Table 3: Parameter estimates for the bandwidth time series model

| Factor | Level | Mb/s |
|---|---|---|
| Intercetp | $\mu$ | 40 |
| *Time* | 0h | 12 |
| | 1h | 13 |
| | 2h | 19 |
| | 3h | 18 |
| | 4h | 15 |
| | 5h | 11 |
| | 6h | 13 |
| | 7h | 9 |
| | 8h | 11 |
| | 9h | 16 |
| | 10h | 14 |
| | 11h | 10 |
| | 12h | 11 |
| | 13h | 5 |
| | 14h | 2 |
| | 15h | 0 |
| | 16h | 5 |
| | 17h | 11 |
| | 18h | 11 |
| | 19h | 10 |
| | 20h | 13 |
| | 21h | 14 |
| | 22h | 8 |
| | 23h | 19 |
| *Weekday* | Mon | 35 |
| | Wed | 2 |
| | Thu | 3 |
| | Tu | 0 |
| | Fri | 4 |
| | Sat | 20 |
| | Sun | 34 |

| F | Level | Mb/s |
|---|---|---|
| $Area^S$ | West US | 69 |
| | North Europe | 30 |
| | US central | 43 |
| | East US | 41 |
| | Australia | 55 |
| | South America | 0 |
| | East Asia | 20 |
| $Area^D$ | West US | 16 |
| | North Europe | 24 |
| | US central | 35 |
| | East US | 60 |
| | Australia | 23 |
| | South America | 0 |
| | East Asia | 40 |
| $CSP^S$ | $C1$ | 7 |
| | $C2$ | 14 |
| | $C3$ | 0 |
| | $C4$ | 24 |
| $CSP^D$ | $C1$ | 0 |
| | $C2$ | 75 |
| | $C3$ | 49 |
| | $C4$ | 78 |
| $CSP^S * CSP^D$ | $C1$ * $C1$ | 0 |
| | $C1$ * $C2$ | 109 |
| | $C1$ * $C3$ | 42 |
| | $C1$ * $C4$ | 308 |
| | $C2$ * $C1$ | 35 |
| | $C2$ * $C2$ | 62 |
| | $C2$ * $C3$ | 6 |
| | $C2$ * $C4$ | 85 |
| | $C3$ * $C1$ | 28 |
| | $C3$ * $C2$ | 30 |
| | $C3$ * $C3$ | 30 |
| | $C3$ * $C4$ | 27 |
| | $C4$ * $C1$ | 71 |
| | $C4$ * $C2$ | 61 |
| | $C4$ * $C3$ | 49 |
| | $C4$ * $C4$ | 0 |

| F | Level | Mb/s |
|---|---|---|
| $Area^S * Area^D$ | West US * West US | 368 |
| | West US * North Europe | 110 |
| | West US * Australia | 109 |
| | West US * Central US | 137 |
| | West US * East US | 123 |
| | West US * South America | 114 |
| | West US * East Asia | 0 |
| | North Europe * West US | 124 |
| | North Europe * North Europe | 337 |
| | North Europe * Australia | 94 |
| | North Europe * Central US | 125 |
| | North Europe * East US | 157 |
| | North Europe * South America | 130 |
| | North Europe * East Asia | 0 |
| | Central US * West US | 164 |
| | Central US * North Europe | 145 |
| | Central US * Australia | 67 |
| | Central US * East US | 182 |
| | Central US * South America | 141 |
| | Central US * Asia | 0 |
| | East US * West US | 169 |
| | East US * North Europe | 177 |
| | East US * Australia | 93 |
| | East US * US | 119 |
| | East US * East US | 365 |
| | East US * South America | 186 |
| | East US * East Asia | 0 |
| | Australia * West US | 57 |
| | Australia * North Europe | 57 |
| | Australia * Australia | 275 |
| | Australia * Central US | 98 |
| | Australia * East US | 57 |
| | Australia * South America | 73 |
| | Australia * East Asia | 0 |
| | South America * West US | 132 |
| | South America * North Europe | 120 |
| | South America * Australia | 122 |
| | South America * Central US | 140 |
| | South America * East US | 136 |
| | South America * South America | 433 |
| | South America * East Asia | 0 |
| | East Asia * West US | 16 |
| | East Asia * North Europe | 0 |
| | East Asia * Australia | 12 |
| | East Asia * Central US | 8 |
| | East Asia * East US | 20 |
| | East Asia * South America | 10 |
| | East Asia * East Asia | 105 |

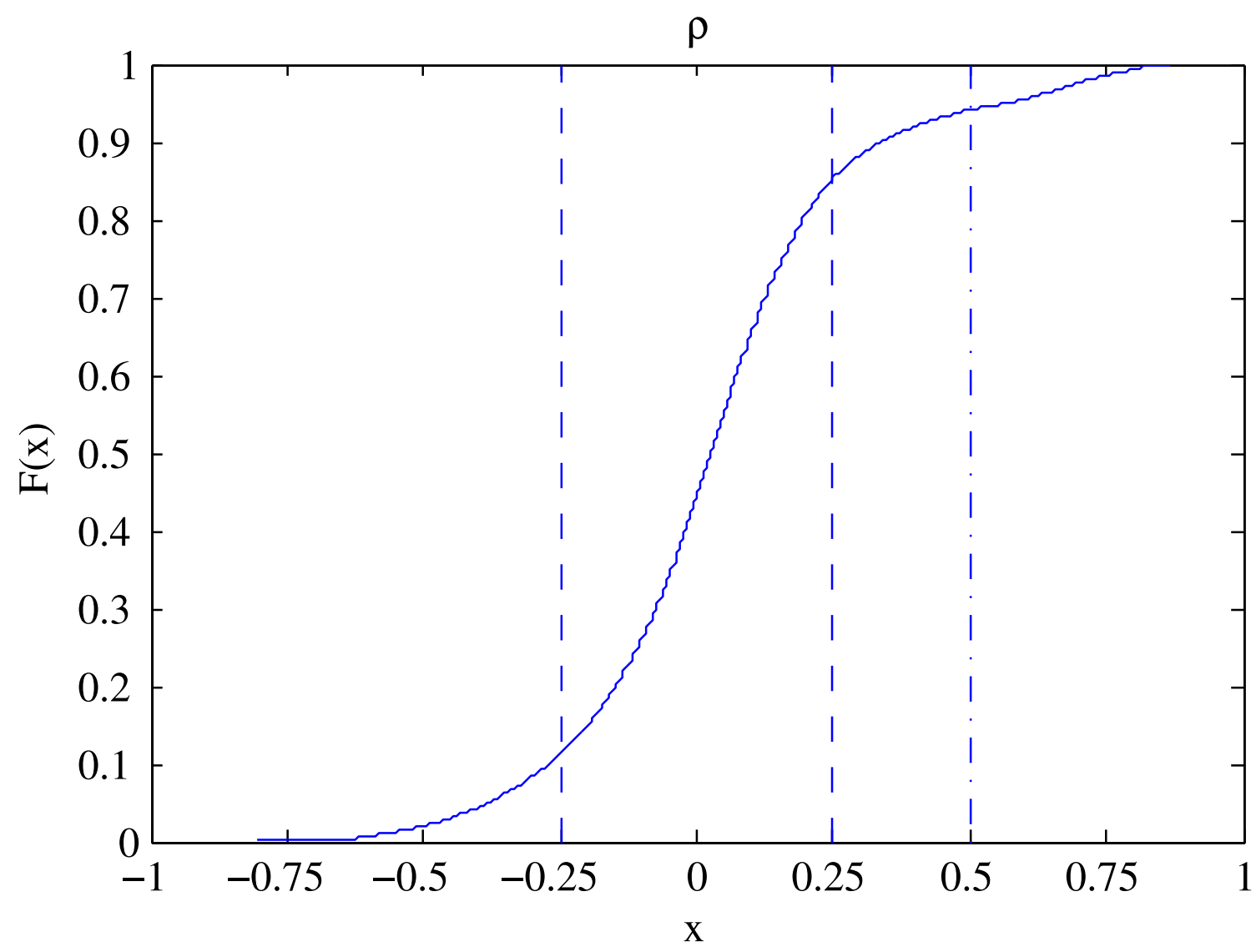

Figure 7: Empirical cumulative distribution function for *rho* of all pair of paths under study

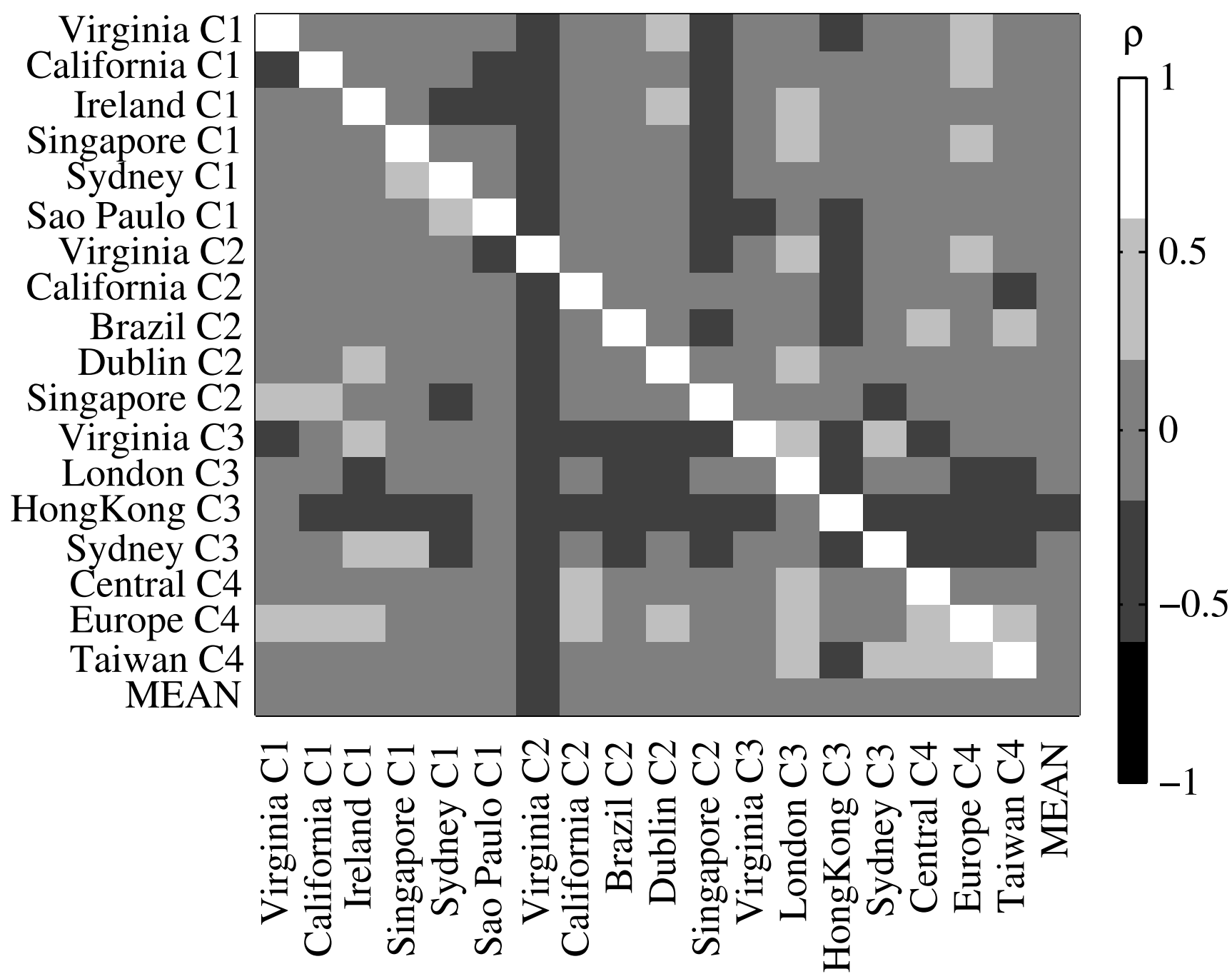

Figure 8: Average *rho* per path (vertical: source data-center, horizontal: destination data-center)

Table 4: ANOVA table with *Area*, *CSP*, *DC* and significant interactions as fixed factors, and the correlation coefficient per pair of paths as response variable

Response variable: Correlation coefficient per pair of paths

| Factor | Sum of Squares | %Total | df | Mean Square | F | $p$-value |
|---|---|---|---|---|---|---|
| $\mu$ | 51.9 | 9.7 | 1 | 51.9 | 752 | 0.00 |
| $Area^S$ | 20.2 | 3.8 | 6 | 3.4 | 49 | 0.00 |
| $Area^{D1}$ | 0.8 | 0.2 | 6 | 0.1 | 2.0 | 0.06 |
| $Area^{D2}$ | 0.9 | 0.2 | 6 | 0.2 | 2.3 | 0.35 |
| $CSP^S$ | 89.0 | 16.6 | 3 | 29.7 | 430 | 0.00 |
| $CSP^{D1}$ | 1.6 | 0.3 | 3 | 0.5 | 7.8 | 0.00 |
| $CSP^{D2}$ | 1.8 | 0.3 | 3 | 0.6 | 8.9 | 0.00 |
| $DC^S$ | 40.9 | 7.6 | 8 | 5.1 | 74 | 0.00 |
| $DC^{D1}$ | 2.6 | 0.5 | 8 | 0.3 | 4.6 | 0.00 |
| $DC^{D2}$ | 2.9 | 0.5 | 8 | 0.4 | 5.3 | 0.00 |
| $Area^S * Area^{D1}$ | 4.6 | 0.9 | 35 | 0.1 | 1.9 | 0.00 |
| $Area^S * Area^{D2}$ | 5.3 | 1.0 | 35 | 0.2 | 2.2 | 0.00 |
| $Area^{D1} * Area^{D2}$ | 8.3 | 1.5 | 35 | 0.2 | 3.4 | 0.00 |
| $CSP^S * CSP^{D1}$ | 9.3 | 1.7 | 9 | 1.0 | 15 | 0.00 |
| $CSP^S * CSP^{D2}$ | 10.5 | 2.0 | 9 | 1.2 | 17 | 0.00 |
| $CSP^{D1} * CSP^{D2}$ | 4.1 | 0.8 | 9 | 0.5 | 6.6 | 0.00 |
| $DC^S * DC^{D1}$ | 40.7 | 7.6 | 227 | 0.2 | 2.6 | 0.00 |
| $DC^S * DC^{D2}$ | 56.6 | 10.6 | 236 | 0.2 | 3.5 | 0.00 |
| $DC^{D1} * DC^{D2}$ | 40.7 | 7.6 | 236 | 0.2 | 2.5 | 0.00 |
| $Area^S * Area^{D1} * Area^{D2}$ | 12.1 | 2.3 | 197 | 0.1 | 0.9 | 0.85 |
| $CSP^S * CSP^S * CSP^S$ | 3.3 | 0.6 | 27 | 0.1 | 1.8 | 0.01 |
| $DC^S * DC^{D1} * DC^{D2}$ | 141.8 | 26.5 | 3807 | 0.0 | 0.5 | 0.00 |
| Error | 0.0 | 0.0 | 0 | | | |
| Total | 535.5 | 100.0 | 4896 | | | |

$\bar{R}^2$=0.99

Table 5: Intercept, $Area^S$ and $CSP^S$ parameter estimates on the correlation parameter model

| Factor | Level | $\rho$ |
|---|---|---|
| Intercept | $\mu$ | 0.10 |
| $Area^S$ | Asia | 0.37 |
| | Brazil | 0.02 |
| | West US | 0.00 |
| | North Europe | 0.09 |
| | Central US | 0.32 |
| | East US | 0.00 |
| | Australia | 0.09 |
| | | |
| $CSP^S$ | $C1$ | 0.27 |
| | $C2$ | 0.26 |
| | $C3$ | 0.38 |
| | $C4$ | 0.00 |